\documentclass[12pt]{article}

\usepackage{amssymb,theorem,amsmath}
\usepackage{epsf}
\usepackage{eucal}

\newcommand{\vect}[1]{\mathbf{#1}}

\let\tsection\section
\renewcommand{\section}{\setcounter{equation}{0}\tsection}

\newcommand{\Ii}{1\!\!1}
\newtheorem{theorem}{Theorem}[section]
\newtheorem{corollary}[theorem]{Corollary}

\newtheorem{proposition}[theorem]{Proposition}

{\theorembodyfont{\rm} 
\newtheorem{remark}[theorem]{Remark} }

\newenvironment{proof}[1][Proof:]{\noindent\textbf{#1} }{\hfill$\blacksquare$}

\def\ds{\displaystyle}
         \def\CC{\mathbb{C}}
    
    \def\NN{\mathbb{N}}
    
    \def\RR{\mathbb{R}}
    \def\ZZ{\mathbb{Z}}
    \def\TT{\mathbb{T}}
    \def\LL{\mathbb{L}}

\let\uacc\u
\def\r{{\bf r}}\def\x{{\bf x}}\def\u{{\bf u}}
\def\y{{\bf y}}\def\z{{\bf z}}
\def\k{{\bf k}}

\def\K{{\cal K}}
\def\etaul{\underline{\eta}}

\begin{document}

\begin{center}\begin{huge}\bf Realizability of point processes \end{huge}
 \vskip20pt
T. Kuna\footnote{Fakult\"at f\"ur Mathematik, Universit\"at Bielefeld,
Postfach 100131, D-33501 Bielefeld.},
J. L. Lebowitz\footnote{Department of Mathematics,
Rutgers University, New Brunswick, NJ 08903.}${}^,$
\footnote{Also Department of Physics, Rutgers.},
and E. R. Speer\footnotemark[2]
\end{center}


\begin{abstract}There are various situations in which it is natural to ask
whether a given collection of $k$ functions,
$\rho_j(\r_1,\ldots,\r_j)$, $j=1,\ldots,k$, defined on a set $X$,
are the first $k$ correlation functions of a point process on $X$.
Here we describe some necessary and sufficient conditions on the
$\rho_j$'s for this to be true.  Our primary examples are $X=\RR^d$,
$X=\ZZ^d$, and $X$ an arbitrary finite set.  In particular, we
extend a result by Ambartzumian and Sukiasian showing realizability
at sufficiently small densities $\rho_1(\r)$.  Typically if any
realizing process exists there will be many (even an uncountable
number); in this case we prove, when $X$ is a finite set, the
existence of a realizing Gibbs measure with $k$ body potentials
which maximizes the entropy among all realizing measures.  We also
investigate in detail a simple example in which a uniform density
$\rho$ and translation invariant $\rho_2$ are specified on $\ZZ$;
there is a gap between our best upper bound on possible values of
$\rho$ and the largest $\rho$ for which realizability can be
established.
\end{abstract}

\section{Introduction\label{intro}}

A {\sl point process} in a set $X$ is a random collection of points in $X$,
whose distribution is described by a probability measure $\mu$ on the set
of all possible point collections.  Here we will take $X$ to be $\RR^d$, a
lattice such as $\ZZ^d$, the torus $\TT^d$, a lattice on the torus, or an
open subset of any of these. We always assume that any bounded subset of
$X$ contains only finitely many points of the collection (this is of course
automatically true if $X$ is a lattice); the collection of points is then
necessarily countable and we will write it as
$\{\vect{x}_1,\vect{x}_2,\ldots\}$, with the understanding that the $\x_i$
are all distinct.

Well known examples of point processes are Gibbs measures for equilibrium
systems of statistical mechanics. The points of the process are then
interpreted as the positions of particles; because the particle
configuration is identified with a subset of physical space, the models
satisfy an exclusion principle: no two particles can occupy the same
position.  Point processes are also used to model phenomena other than
those of statistical mechanics, such as trains of neural spikes
\cite{Mitra1,Mitra2}, departure times from queues \cite{queues}, and
positions of stars in galaxies \cite{galaxies}.

In many of these cases the quantities of primary interest, partly because
they are the ones most accessible to experiment, are the low order
correlations, such as the one particle density $\rho_1({\bf r}_1)$ and the
pair density $\rho_2({\bf r}_1,{\bf r}_2)$.  These may be defined in terms
of expectations (averages), with respect to the measure $\mu$, of products
of the (random) {\sl empirical field} $\eta$ describing the process. For
continuum systems, i.e., when $X$ is $\RR^d$ or an open subset of $\RR^d$,
$\eta$ is defined by
 \begin{equation}
\eta({\bf r}) = \sum_i \delta({\bf r}-{\bf x}_i),\label{eta}
 \end{equation}
  where $\delta$ is the Dirac delta function and the $\x_i$'s are as above
the (random) positions of the points of the process.  Then, with
$\langle\cdot\rangle$ denoting expectation with respect to $\mu$,
 \begin{eqnarray}
  \rho_1({\bf r}_1) &=& \langle \eta({\bf r}_1) \rangle  , \label{defrho1}\\
  \rho_2({\bf r}_1,{\bf r}_2) &=&
  \langle \eta({\bf r}_1)\eta({\bf r}_2)\rangle
     -\rho_1(\r_1)\delta(\r_1-\r_2) ,\label{defrho2}
 \end{eqnarray}
 and so on; in general,
 \begin{equation}
\rho_n(\r_1,\ldots,\r_n)=\left\langle\sum_{i_1\ne i_2\ne\cdots\ne i_n}
   \prod_{k=1}^n\delta(\r_{k}-\x_{i_k})\right\rangle.\label{defrhon}
 \end{equation}
 Equation (\ref{defrhon}) defines $\rho_n$ as a measure, and we will
always assume that this measure assigns finite mass to any bounded
set in $\RR^{dn}$ (which means that the number of particles in a
bounded set in $\RR^d$ is a random variable with finite moments up
to order $n$).  In many cases this measure is absolutely continuous
with respect to Lebesgue measure and we identify
$\rho_n(\r_1,\ldots,\r_n)$ with its density, i.e.,
$\rho_n(\r_1,\ldots,\r_n)d\r_1\ldots d\r_n$ is the probability of
finding a particle in the infinitesimal box $d\r_i$ at $\r_i$ for
$i=1,\ldots ,n$. We will assume that, whenever possible, $\rho_n$ is
extended by continuity to be defined at points where two of its
arguments coincide.  The $\rho_n$, here and in the lattice case
discussed in the next paragraph, are often referred to as the
$n$-particle distribution functions or correlation functions.

 When $X$ is discrete, a finite set or a lattice, we will also use the
notation (\ref{eta}) and the definitions (\ref{defrho1})--(\ref{defrhon}),
interpreting $\delta({\bf r}-{\bf x}_i)$ as the Kronecker delta
$\delta_{{\bf r},{\bf x}_i}$, so that $\eta(\r)$ has value 0 or 1,
depending on whether the site $\r\in X$ is empty or occupied.  Note that if
$n\ge2$ then $\rho_n(\r_1,\ldots,\r_n)$, ${\bf r}_i \in X$, vanishes
whenever $\r_i=\r_j$ for some $i\ne j$, and that for distinct sites
$\r_1,\ldots,\r_n$, $\rho_n(\r_1,\ldots,\r_n)$ is the probability of having
particles at these sites.

In this paper we shall study the following {\sl realizability problem:}
given functions $\rho_1({\bf r}_1)$, $\rho_2({\bf r}_1, {\bf r}_2)$,
$\ldots$, $\rho_k({\bf r}_1,\ldots,{\bf r}_k)$, nonnegative and symmetric,
does there exist a point process for which these are the correlations?
Since only a finite number of correlations are prescribed, the problem may
be regarded as an infinite dimensional version of the standard truncated
moment problem \cite{trunc}.  The full realizability problem, in which all
the correlations $\rho_j$, $j=1,2,\ldots$ are given, was studied by
A.~Lenard \cite{Lenard}.

Realizability, and the related question of fully describing the
realizing process, are long standing problems in the classical
theory of fluids \cite{Percus}, \cite{GarrodPercus}, and
\cite{Hansen}, recently revived by Torquato, Stillinger, et
al.~\cite{ST,ST2}. This problem has been also investigated in the
context of the quantum system see e.g. \cite{GarrodPercus},
\cite{Kummer}, and \cite{ColeYuk}. An important ingredient in that
theory is the introduction of various approximation schemes for
computing $\rho_2(\r_1,\r_2)$, such as the Percus-Yevick and
hyper-netted chain approximations \cite{Hansen}. It is then of
primary interest to determine whether or not the resulting functions
$(\rho_1,\rho_2)$ in fact correspond to any point process, that is,
are in some sense internally consistent. If so  they can provide
rigorous bounds for the entropy of the system under consideration. A
novel application of the realizability problem to the determination
of the maximal density of sphere packing in high dimensions is
discussed in \cite{ST3}.

Applications of the problem of describing a point process from
its low order correlations occur in other contexts, for example, in
the study of neural spikes \cite{Mitra1, Mitra2}.  In this and other
physical situations it is natural to consider a closely related
problem in which the $\rho_j$, $j=1,2,\ldots,k$, are
specified only on part of the domain $X$; for example, on the lattice we
might only specify the nearest neighbor correlations.  See \cite{Kanter}
for a similar problem in error correcting codes.  We will not
consider this case further here, except for some comments at the end
of section 6.

An important special case is that  in which $X$ is  $\RR^d$,
$\ZZ^d$, or a periodic version of one of these (a torus), and the point
process is translation invariant.  The specified correlation functions
will then also be translation invariant and may be written in the form
 \begin{eqnarray}
  \rho_1({\bf r}_1) &=& \rho,\label{rho}\\
 \rho_j({\bf r}_1,\ldots,{\bf r}_j)
  &=&\rho^jg_j({\bf r}_2-{\bf r}_1,\ldots,{\bf r}_j-{\bf r}_1),
 \qquad j=2,\ldots,k.\label{g}
 \end{eqnarray}
 As we often work with $k=2$, we write $g(\r)\equiv g_2(\r)$.
 We will often state our arguments and results in the translation invariant
case, but these may frequently be extended to the more general situation;
when we do not impose translation invariance we will use a notation
similar to (\ref{g}):
 \begin{equation}
 \rho_j({\bf r}_1,\ldots,{\bf r}_j)
  = \prod_{i=1}^j\rho_i(\r_i)
    \;G_j({\bf r}_1,\r_2,\ldots,{\bf r}_j),
 \qquad j=2,\ldots,k.\label{ntig}
 \end{equation}

We now make some general remarks in order to put the realizability
problem in context.  First, we observe that if the correlations
(\ref{rho})--(\ref{g}) can be realized for some density $\rho$, then
they can also be realized, for the same functions $g_2,\ldots,g_k$,
for any $\rho'$ with $0\le\rho'<\rho$ \cite{AS}. To see this, note
that if $\eta_0(\r)=\sum \delta({\bf r}-{\bf x}_i)$ is the empirical
field with density $\rho$, then $\eta(\r)=\sum Q_i\delta({\bf
r}-{\bf x}_i)$, where the $Q_i$ are independent, identically
distributed Bernoulli random variables with expectation
$\rho'/\rho$, is a field with density $\rho'$ having the same value
for all $g_j$'s.  In other words, the new measure is constructed by
independently choosing to delete or retain each point in a
configuration, keeping a point with probability $\rho'/\rho$. (We
will refer to this construction as {\sl thinning}.)  In this light
it is thus natural to pose the realizability problem in the
following form: given the $g_j$, $j=2,\ldots,k$, what is the least
upper bound $\bar\rho$ of the densities for which they can be
realized?  It is of course possible in the continuum case to have
$\bar\rho=\infty$; for example, if $g_j=1$ for $j\le2\le k$ then for
any density $\rho>0$ a Poisson process realizes the correlations.
For the lattice systems considered here, on the other hand, we
always have $\bar\rho\le1$.

Lacking a full answer to this question, one may of course ask rather for
upper and lower bounds on $\bar\rho$.  A lower bound $\bar\rho\ge\rho_0$
may be obtained by the construction of a process at density $\rho_0$; we
discuss such constructions in sections~\ref{lowrho}--\ref{subsecLee} and
show that $\bar\rho>0$ under reasonable restrictions on the $g_j$ in
(\ref{g}).  Upper bounds on $\bar\rho$ may be obtained from necessary
conditions for realizability, some of which are described in
section~\ref{secneccond}.

Beyond the question of realizability one may ask about the number or more
generally about the types of measures which give rise to a specified set of
correlations $\rho_j$, $j=1,\ldots,k$.  A natural question in the theory of
fluids, for example, is whether any of these measures are Gibbsian for
interactions in a particular class; for example, given $\rho_1$ and
$\rho_2$, is there a Gibbs measure realizing these correlations which
involves only pair interactions? This question is considered in
section~\ref{gibbs}, where we discuss the nature of the realizing measure
$\mu$ which maximizes the Gibbs-Shannon entropy.

 The outline of the rest of this paper is as follows.  In
section~\ref{secneccond}, we discuss some necessary conditions for
realizability.  (In a separate paper \cite{KLSnecsuff} we will present
general necessary and sufficient conditions.)  Sections~\ref{lowrho},
\ref{triplet}, and \ref{subsecLee} cover proofs of realizability: in
section~\ref{lowrho} we prove a theorem, a generalization of one proven by
R.~V.~Ambartzumian and H.~S.~Sukiasian \cite{AS}, showing that if $g-1$ is
absolutely integrable and $g$ satisfies a certain stability condition then
the pair $(\rho,g)$ is realizable for sufficiently small $\rho$, with
explicitly given higher correlations.  In section~\ref{triplet} we show
that the construction can be extended to the case in which the third
correlation function $g_3$ is also specified, showing in particular that
the realization determined by $(\rho,g)$ alone is not unique; in fact since
$g_3$ can take an uncountable number of values there are at low values of
$\rho$ an uncountable number of measures realizing $(\rho,g)$.  We note
also possible extensions to higher order $g_j$.  In section~\ref{subsecLee}
we give a variant construction for lattice systems, based on the Lee-Yang
theorem \cite{LY,Ruelle}.  In section~\ref{gibbs} we show that a problem with
specified $(\rho,g)$, on a finite set, e.g., a periodic lattice, may be
realized by a Gibbs measure with just one- and two-particle potentials
whenever $\rho<\bar\rho$.  We make some concluding remarks in
section~\ref{conclude}, and in the appendices discuss in some detail an
illustrative one-dimensional example and give some technical proofs.

\section{Necessary conditions for realizability\label{secneccond}}

Clearly, from (\ref{defrhon}), realizability requires that
 \begin{equation}
\label{neccond1} \rho_j(\vect{r}_1,\ldots,\vect{r}_j) \geq 0,\qquad
j=1,\ldots,k.
 \end{equation}
 We also know that the covariance matrix of the field $\eta(\r)$,
 \begin{eqnarray}
  S(\r_1,\r_2)&=&\langle\eta(\r_1)\eta(\r_2)\rangle
        -\langle\eta(\r_1\rangle\langle\eta(\r_2)\rangle\label{cov}
        \\
        &=&\rho_2(\r_1,\r_2) + \rho_1(\r_1) \delta(\r_1-\r_2)
        -\rho_1(\r_1)\rho_1(\r_2), \nonumber
 \end{eqnarray}
 must be positive semi-definite, i.e., for all functions $\varphi$ with bounded
 support,
 \begin{equation}
 \label{covform}
 \int \int \varphi(\r_1) \bar{\varphi}(\r_2) S(\r_1,\r_2)\, d\r_1 d\r_2
 \geq 0.
 \end{equation}
 If we take $\varphi(\r)=\Ii_\Lambda e^{i\k\r}$ for
$\Lambda$ a bounded subset of $X$ then (\ref{covform}) becomes
\begin{equation}
  \int_\Lambda\rho_1(\r_1)\,d\r_1 +  \int_{\Lambda}
\int_{\Lambda}   e^{i {\bf k}\cdot (\r_1-\r_2)}
\left[\rho_2(\r_1,\r_2) -\rho_1(\r_1)\rho_1(\r_2) \right]
d{\r_1}d\r_2 \geq
   0;   \label{FTl}\\
 \end{equation}
 conversely, if (\ref{FTl}) holds for all ${\bf k}\in\RR^d$  and all
 $\Lambda$, then (\ref{covform}) holds for all $\varphi$.

 In the translation invariant case (\ref{FTl}) is equivalent to the
nonnegativity of the infinite volume structure function $\hat{S}({\bf k})$:
 \begin{equation}
  \hat{S}({\bf k}) \equiv \rho + \rho^2 \int_{\mathbb{R}^d}
   e^{i {\bf k}\cdot {\bf r}} \left[g({\bf r}) -1 \right] d{\r} \geq
   0.
  \label{FT}\\
 \end{equation}
 Here we assume
$\int_{\mathbb{R}^d} \left|g({\bf r}) -1 \right| d{\r} < \infty$; otherwise
(\ref{FT}) holds in the sense of generalized functions, cf. \cite{genfct}.
There are corresponding conditions on the torus $\TT^d$, the lattice
$\ZZ^d$, and the periodic lattice. If equality holds in (\ref{FT}) for some
${\bf k}$ then clearly  $\rho$ is maximal: $\rho=\bar{\rho}$.

 We note also a necessary condition due to Yamada \cite{Yamada}: if
$N_\Lambda$ denotes the number of of particles in a region
$\Lambda\subset X$, and if $\theta$ is the fractional part of the mean of
$N_\Lambda$, so that $\langle N_\Lambda \rangle = k+ \theta$ with
$k=0,1,\ldots$ and $0\le\theta<1$, then the variance $V_\Lambda$ of
$N_\Lambda$,
 \begin{eqnarray}
V_\Lambda &\equiv& \int_\Lambda \int_\Lambda S(\r_1,\r_2) d\r_1
d\r_2 \nonumber \\
&=&    \int_\Lambda\rho_1(\r_1)\,d\r_1 +
   \int_\Lambda\int_\Lambda
      \left[\rho_2({\bf r}_1,{\bf r}_2) -\rho_1(\r_1)\rho_1(\r_2) \right]
    d{\bf r}_1 d{\bf r}_2,
  \label{var}
 \end{eqnarray}
 must satisfy
 \begin{equation}\label{Yamada}
V_\Lambda\ge\theta(1-\theta),
 \end{equation}
 because $N_\Lambda$ is an integer:
 \begin{equation} V_\Lambda = \langle(N_\Lambda-k)^2 \rangle - \theta^2 =
\theta(1-\theta)+\Big\langle \bigl((N_\Lambda-k)(N_\Lambda-k-1)\bigr)
\Big\rangle \ge\theta(1-\theta).  \label{theta} \end{equation}

The above necessary conditions all follow from the more general
conditions that we prove in \cite{KLSnecsuff}. In summary these say that,
given any functions $f_2({\bf r}_1, {\bf r}_2)$, $f_1({\bf r})$ and
 constant $f_0$ such that, for any $n$ points ${\bf
r}_1,\ldots,{\bf r}_n$ in $X$, $\sum_{i \neq j} f_2({\bf r}_i,{\bf
r}_j) + \sum_i f_1({\bf r}_i) +f_0 \geq 0$, we must have
\begin{equation}
   \mathop{\int\!\!\int}_{_{\!\!\!\scriptstyle \Lambda\times\Lambda}}
   \rho_2({\bf r}_1,{\bf r}_2) f_2({\bf r}_1,{\bf r}_2) d{\bf r}_1 d{\bf r}_2
 + \int_{\Lambda} \rho_1(\r)f_1({\bf r}) d{\bf r} +f_0
    \geq 0, \label{all}
 \end{equation}
 for all $\Lambda \subset \mathbb{R}^d$. We prove in fact that in the case
$k=2$, i.e. for the case that only $\rho_1$ and $\rho_2$ are given,
(\ref{all}) is also a sufficient condition for realizability under some
mild assumptions on the point process.

We remark that in the case $k=2$ all restrictions on $\rho$ and $g$ beyond
those arising from nonnegativity of $\rho$ and of the covariance matrix $S$
of (\ref{cov}) are due to the fact that we want $\eta(\r)$ to be a point
process, since we can always find a Gaussian process realizing any
$\rho_1$, $\rho_2$ with $S>0$ \cite{Gaussian}.

We also note that for $g(\r)\leq 1$ one has
 \begin{equation}
\hat{S}({\bf k}) \geq \hat{S}({\bf 0}) = \lim_{\Lambda \rightarrow
\infty } \frac{1}{|\Lambda|} V_\Lambda .
 \end{equation}
 Hence equality in (\ref{FT}) implies that the variance $V_\Lambda$ is
growing slower than the volume. Processes with this property are called
{\sl superhomogeneous} and are of independent interest, see
\cite{superhom,ST,GLS}. As noted above, superhomogeneity can hold, for a
given $g(\r)$, only at the maximal density $\rho=\bar{\rho}$.

\section{Realizability for small density\label{lowrho}}

In this section we show the realizability of a given translation invariant
$\rho$ and $g(\r)$, $\r \in \RR^d$, for sufficiently small $\rho$. Our
arguments extend immediately to the lattice $\ZZ^d$ and to the torus, as
well as to non-translation invariant $\rho_1(\r)$, $\rho_2(\r_1,\r_2)$.
Our results are an extension of those given by R.~V.~Ambartzumian and
H.~S.~Sukiasian \cite{AS}, and are based closely on the key idea of that
paper: given $\rho$ and $g(\r)$ satisfying suitable conditions,
one proves the existence of a translation
invariant process for which the correlation functions $\rho_n$,
$n=1,2,3,\ldots$, are given by
\begin{eqnarray}
\rho_n(\r_1,\ldots,\r_n) &=& \rho^n \prod_{1 \leq i < j \leq n}
g(\r_i - \r_j), \label{armansatz}
\end{eqnarray}
 and which therefore solves the realization problem for $\rho$ and $g$.
The ansatz (\ref{armansatz}) for the dependence on $\rho$ and $g$
of the higher order
correlations (which determines the point process
uniquely) is arguably the simplest one possible. It corresponds, for $n=3$,
to the well-known Kirkwood superposition approximation in the theory of
equilibrium fluids \cite{Hansen}. Here, however, we are not talking of an
approximation to a particular given point process but rather of
constructing a realizing process or measure $\mu$ whose correlations have
the form (\ref{armansatz}).

 To find a point process corresponding to (\ref{armansatz}), Ambartzumian
and Sukiasian used the inclusion-exclusion principle which, for any
point process, relates the correlation functions $\rho_n$ to the
probability densities $p_n^{\Lambda}(\r_1,\ldots,\r_n)$ for finding
exactly $n$ particles, with positions $\r_1,\ldots,\r_n$, in a region
$\Lambda \subset \RR^d$:
 \begin{eqnarray} \label{inclus}
  p^\Lambda_n(\r_1,\ldots,\r_n)\nonumber\hskip-50pt&&\\
&=& \sum_{k=0}^\infty \frac{(-1)^k }{k!} \int_{\Lambda^k}
\rho_{n+k}(\r_1,\ldots,\r_n,\x_1,\ldots,\x_k) \, d\x_1\ldots d\x_k.
\end{eqnarray}
Inserting the ansatz (\ref{armansatz}) in (\ref{inclus}), one
finds an expression for the proposed densities:
\begin{eqnarray}
p^\Lambda_n(\r_1,\ldots,\r_n)
  &\equiv&  \rho^n \prod_{1 \leq i < j \leq n}
   g(\r_j - \r_i)
\nonumber \\
&& \hskip-80pt
  \times \sum_{k=0}^\infty \frac{(-\rho)^k }{k!} \int_{\Lambda^k}
\prod_{1\leq i < j \leq k} g(\x_j - \x_i) \prod_{\stackrel{
\scriptstyle i=1,\ldots,n}{j=1,\ldots ,k}} g(\x_j - \r_i)
\,d\x_1\ldots d\x_k. \label{armmeasure}
 \end{eqnarray}
 It remains to verify that (\ref{armmeasure}) in fact defines the
probability densities of a point process.

First, note that the quantities $p^{\Lambda}_n(\r_1,...,\r_n)$ are well
defined by (\ref{armmeasure}) for any value of $\rho$ whenever there is,
for every region $\Lambda$, a constant $M_\Lambda$ such that
(\ref{armansatz}) satisfies
 \begin{equation}
|\rho_n(\r_1,\ldots,\r_n)| \leq M_\Lambda^n, \label{clam}
 \end{equation}
 for $\r_1,\ldots,\r_n$ in $\Lambda$.  The condition (\ref{clam}) is easily
verified for many $g$ (see, e.g., Theorem~\ref{ThArmenianCluster} below).
The remaining problem is to prove that the $p^\Lambda_n$ are all
nonnegative.  If this is done, then in each region $\Lambda$ the collection
$p^\Lambda_n$, $n=1,2,\ldots$, determines a measure $\mu^\Lambda$ defining
a point process; if $\Lambda\subset\Lambda'$ then $\mu^\Lambda$ and
$\mu^{\Lambda'}$ are compatible and by general arguments (Kolmogorov's
projective limit theorem) there exists an infinite volume realizing measure
$\mu$.

Ambartzumian and Sukiasian considered only the case $g(\r)\leq 1$,
$r\in\RR^d$. For this case they constructed a cluster expansion of the
Penrose-Ruelle type and obtained inequalities of the
Groeneveld-Lieb-Penrose type to show nonnegativity of each term in a
reorganized expansion of the $p_n^\Lambda$.  In order to extend their
result to $g$'s which can be bigger than one, when the cluster expansion is
no longer positive term by term, we need to use a different approach.
Recall the definition
 of the standard grand canonical partition function, in the region
$\Lambda$, of a particle system with fugacity $z$, one-particle potential
$V^{(1)}(\y)$, pair potential $V(\r)$, and inverse temperature $\beta=1$
\cite{Ruelle}:
 \begin{eqnarray}
    \Xi_\Lambda(z,V^{(1)},V)
    &=& \sum_{k=0}^{\infty} \frac{z^k}{k!}\int_{\Lambda}\cdots\int_\Lambda
   \nonumber\\
  && \hskip-60pt \label{partfunct}
   \times\exp\biggl\{-\biggl[\sum_{1\le i\le k} V^{(1)} (\y_i)
    +  \sum_{1\le i<j\le k} V(\y_i-\y_j)\biggr]\biggr\}\, d\y_1\cdots d\y_k.
 \end{eqnarray}
  Then (\ref{armmeasure}) takes the form
\begin{equation} \label{GCPF}
  p_n^\Lambda(\r_1,\ldots,\r_n)
    = \left[\rho^n \prod_{1 \leq i < j \leq n}
     g(\r_i - \r_j)\right] \Xi_\Lambda(-\rho,V^{(1)},V),
\end{equation}
 with $V(\r)=-\log (g(\r))$ and
$V^{(1)}(\y)\bigl(=V^{(1)}(\y;\r_1,\ldots,\r_n)\bigr)
=\sum_{i=1}^n V({\bf
\y}-\r_i)$.
Note that in (\ref{GCPF}) the one-particle potential $V^{(1)}$, and hence
also the partition function $\Xi_\Lambda(z,V^{(1)},V)$, is different for
each $ p_n^\Lambda(\r_1,\ldots,\r_n)$, depending explicitly on $n$ and on
the particle positions $\r_1,...,\r_n$.  The condition (\ref{clam}) implies
that $\Xi_\Lambda(z,V^{(1)},V)$ is an entire function of $z$.

Suppose now that $\log\Xi_\Lambda(z,V^{(1)},V)$ is
analytic in $z$ in some domain $\Omega$ containing the origin; then
$\Xi_\Lambda(z,V^{(1)},V)$ can not vanish in $\Omega$.  In particular, if
$(a,b)$ is the largest interval on the real axis which contains the origin
and is contained in $\Omega$, then $\Xi_\Lambda(z,V^{(1)},V)>0$ for
$a<z<b$, since $\Xi_\Lambda(0,V^{(1)},V)=1$.  We will apply this
observation by finding a domain $\Omega$---a disk centered at the origin,
of radius $R$---such that for all $\Lambda$, all $n$, and all
$\r_1,\ldots,\r_n$, $\log\Xi_\Lambda(z,V^{(1)},V)$ is analytic in $\Omega$;
then since $g(\r)\ge0$, all $p^\Lambda_n(\r_1,\ldots,\r_n)$ will be
nonnegative for $0\le\rho\le R$.

These considerations lead to our main result.  We write
 \begin{equation}
   C(g)\equiv\int_{\RR^d}|g(\x)-1| d\x.
 \end{equation}

 \begin{theorem} \label{ThArmenianCluster} Let $g$ be a non-negative even
function on $\RR^d$, and suppose that (i)~$C(g)<\infty$ and (ii)~there
exists a constant $b$, $1\le b<\infty$, such that for all $n\ge1$,
 \begin{equation}\label{b}
\prod_{i=1}^{n}g(\x_0-\x_i) \leq b
 \end{equation}
  whenever
$\x_0,\x_1,\ldots,\x_n$ satisfy $\prod_{i<j} g(\x_i - \x_j) >0$.  Then
(\ref{clam}) is satisfied, and $(\rho,g)$ is realizable, for all $\rho$
satisfying
\begin{equation}
\label{ThArmeniancluster Req} 0 \leq \rho \leq \left( eb C(g)\right)^{-1}.
\end{equation}
 \end{theorem}

For completeness we state the analogous result on the lattice.

\begin{theorem} \label{armenian-lat} Let $g$ be an even non-negative
function on $\mathbb{Z}^d$, and suppose that
$C(g):=\sum_{x\in \mathbb{Z}^d} |g(x)-1| < \infty$.  Let $b$ be a constant,
with $1\le b<\infty$, such that $\prod_{i=1}^n g(x_i)\le b$ whenever
$x_1,\ldots,x_n$ satisfy $\prod_{i<j} g(x_i -x_j) >0$. Then $(\rho,g)$ is
realizable for all $\rho$ satisfying $0\leq \rho\leq (ebC(g))^{-1}$.
\end{theorem}

\begin{remark} \label{armremark} (a) The fact that (\ref{clam}) holds under
hypothesis (ii) of the theorem, with constant
$M_\Lambda=\rho b^{1/2}$ independent of $\Lambda$, is immediate. In the
language of statistical mechanics, this says that the interaction $V$ is
{\sl stable}.

\noindent (b) If $g\leq 1$ then hypothesis (ii) holds
with $b=1$, and we recover the result of \cite{AS}.

\noindent (c) Hypothesis (ii) also holds if there exists (I)~a $D>0$
such that $g(\r)=0$ when $|\r| \leq D$, and (II)~a nonnegative decreasing
function $\psi$ on $[D,\infty)$, satisfying
$\int_0^\infty t^{d-1}\psi(t)\,dt<\infty$, such that
$(g(\r)-1)\leq \psi(|\r|)$ \cite{Ruelle}.  In the
language of statistical mechanics, (I) says that $V(\r)$ has a {\sl hard
core}; (II) says that $V(\r)$ is  {\sl lower regular} \cite{Ruelle2}.  In
this case one easily obtains an explicit possible value for the constant $b$.

\noindent (d) Note that, despite the use of results from equilibrium
systems, there is no reason to expect the realizing measure $\mu$
giving rise to the $\rho_n$ of (\ref{armansatz}) to be a
Gibbs measure with pair potential (unless $g(\r)\equiv1$, in which case
$V(\r) = 0$ and $\mu$ corresponds to a Poisson process).

\end{remark}

\begin{proof}[Proof of Theorem \ref{ThArmenianCluster}:]
 Denote by $k^\Lambda_m$ the $m^{\rm th}$ correlation function for a grand
canonical ensemble in $\Lambda$ with pair potential $V$, one-particle
potential $V^{(1)}$, inverse temperature $\beta=1$, and activity $z$; as
for the partition function (\ref{partfunct}) for this system,
these correlation functions depend through
$V^{(1)}$ on $\r_1,\ldots,\r_n$. By the above remarks it suffices to
establish analyticity of $k^\Lambda_1$ in a disk $|z|< R$ with
$R=\left(ebC(g)\right)^{-1}$, because from
 \begin{equation}
\frac{d}{dz} \log \left( \Xi_\Lambda(z,V^{(1)},V)
\right) = \frac{1}{z} \int_\Lambda k^\Lambda_1(\x) \,d\x
 \end{equation}
 it follows that $\log\Xi_\Lambda(z,V^{(1)},V)$ is also analytic in this disk.

To establish the analyticity of $k^\Lambda_1$ we in fact show
analyticity of all $k^\Lambda_m$; we proceed as in the classical
proof, following in particular section~4.2 of \cite{Ruelle}.  In this
 proof the Kirkwood-Salsburg equations for the correlation
functions are written in an appropriately chosen Banach space in the
form $k^\Lambda=z\psi+z\K k^\Lambda$ for some operator $\K$ and fixed
vector $\psi$. One shows that $\|z\K\|<1$ when $|z|<R$, so that $I-z\K$ is
then invertible via a power series in $z$, and  a
unique solution, analytic in $z$, exists. The primary change in
the proof required in our case is that one must introduce a
dependence of the operator $\K$ on the sites $\r_1,\ldots,\r_n$. We
leave the details to appendix~\ref{armproof}, which is probably best
read with \cite{Ruelle} in hand.  \end{proof}

We next state a generalization of Theorem~\ref{ThArmenianCluster} to
non-translation invariant systems; the proof is omitted.  Let
$X\subset\RR^d$ be open, and recall the notation
$\rho_2(\x,\y)=\rho_1(\x)\rho_1(\y)G_2(\x,\y)$ of (\ref{ntig}).

\begin{theorem} Let $\rho_1$ and $G_2$ be non-negative functions on $X$ and
$X\times X$, respectively, with $G_2$ symmetric, and suppose that there
exists a constant $b$, with $1\le b<\infty$, such that for all $n\ge1$,
$\prod_{i=1}^{n}G_2(\x_0,\x_i) \leq b$ whenever
$\x_0,\x_1,\ldots,\x_n \in X$ satisfy $\prod_{i<j}\rho_2(\x_i-\x_j)>0$.
Then the pair $(\rho_1,\rho_2)$ is realizable if
\begin{equation}
eb\  \sup_{\x\in X}\left(\int_X |G_2(\x,\y)-1| \rho_1(\y)d\y\right) \leq 1.
\end{equation}
\end{theorem}

%

\subsection{Decay of correlations \label{decay}}

We are interested in the decay of the truncated correlation functions $u_k$
for the realizing measure specified by (\ref{armansatz}), defined
recursively by \cite{Ruelle}
\begin{eqnarray}
\rho_n(\r_1,\ldots,\r_n) = \sum_{k=1}^n \sum_{\{I_1,\ldots,I_k \}
\in \mathcal{P}_k(n)} \prod_{j=1}^k u_{|I_j|}((\r_i)_{i \in I_j}),
\end{eqnarray}
 where $\mathcal{P}_k(n)$ denotes the set of all partitions of
$\{1,\ldots,n \}$ in $k$ disjoint sets. We consider only the case in which
$X=\RR^d$ and $\rho_1$ and $\rho_2$ are translation invariant; then
$u_1(\r_1)=\rho$ and $u_2(\r_1,\r_2)=\rho^2[g(\r_1-\r_2)-1]$.  For the
correlation functions (\ref{armansatz}) the corresponding truncated
correlation functions have the form
 \begin{equation} \label{graphs}
u_n(\r_1,\ldots,\r_n) = \rho^n \sum_{G
\in \mathcal{G}_c(n)} \prod_{\{i,j\} \in G}
\big(g(\r_i-\r_j)-1\big),
 \end{equation}
 with $\mathcal{G}_c(n)$ the set of all connected subgraphs of the complete
 graph with vertex set $\{1,2,\ldots,n\}$.

Let $\mathcal{T}(n)$ denote the set of all undirected trees on
$\{1,\ldots,n\}$. Then from (\ref{graphs}) and an estimate of Penrose
\cite{penrose},
 \begin{equation}
 |u_n(\r_1,\ldots,\r_n)|
\leq \rho^nb^{n-2} \sum_{T \in \mathcal{T}(n)}
\prod_{\{i,j\} \in T} \big|g(\r_i-\r_j)-1\big|,
 \end{equation}
where $b$ is defined as in Theorem~\ref{ThArmenianCluster}. Using
$|\mathcal{T}(n)|=n^{n-2}$ we then obtain the $L^1$ decay  property
 \begin{equation}
 \int_{X^n} \big|u_{n+1}(\r_0,\r_1,\ldots,\r_n) \big| \,d^n \r
  \leq \rho^{n+1}((n+1)b)^{n-1}C(g)^n.
  \end{equation}
 Compare Theorem 4.4.8 of \cite{Ruelle}.

 One may also  establish a  pointwise decay bound:  if
$|g(\r)-1|$ decays polynomially or exponentially, then
$u_n(\r_1,\ldots,\r_n)$ also decays polynomially or exponentially,
respectively, with $\max_{1\leq i < j \leq n} |\r_i -\r_j|$. For example,
if $|g(\r)-1| \leq D_1 e^{-D_2|\r|}$ for some $D_1,D_2>0$ and all
$\r$, then
\begin{equation}
|u_n(\r_1,\ldots,\r_n)|
   \leq (nb)^{n-2}D_1^{n-1}\rho^n e^{-D_2 L},
\end{equation}
where $L$ is the minimal length of a tree connecting all points
$\r_1,\ldots,\r_n$ and the length of a tree $T$ is
$\sum_{\{i,j\}\in T}|\r_i-\r_j|$.

This decay implies that the realizing measure is mixing and
therefore ergodic.

\section{Triplet correlation function \label{triplet}}

We now consider briefly the application of the ideas of
section~\ref{lowrho} to the realization problem under the specification of
$\rho_1,\ldots,\rho_k$ for $k>2$. For simplicity we discuss only the case
$k=3$ and restrict our considerations to translation invariant correlations
in $\RR^d$ which have densities with respect to Lebesgue measure.  Other
cases can be treated analogously.  We adopt the notation
 \begin{eqnarray}
\rho_3(\x,\y,\z) &=& \rho^3g_3(\y-\x,\z-\x)\nonumber\\
  &=& \rho^3 g(\y-\x) g(\z-\x)g(\z-\y)\tilde{g}_3(\y-\x,\z-\x);
 \end{eqnarray}
 the first equation here is just (\ref{g}), and the second is justified by
the fact that, from the definition (\ref{defrhon}), $\rho_2(\x,\y)$ cannot
vanish on a set $S$ of positive (Lebesgue) measure unless $\rho_3(\x,\y,\z)$
vanishes, for almost all $\z$, if ($\x,\y)\in S$.

In analogy with (\ref{armansatz}) we make now the ansatz
\begin{eqnarray}
\label{armansatztriplet} \rho_n(\r_1,\ldots,\r_n) &:=& \rho^n
\prod_{1\leq i < j \leq n} g(\r_i -\r_j) \prod_{1\leq i < j < k \leq
n} \tilde{g}_3(\r_k-\r_i,\r_j-\r_i)
\end{eqnarray}
for the higher correlation functions ($n \ge4$). As before, the probability
densities $p_n^\Lambda$ for the point process defined by
(\ref{armansatztriplet}) can be written in terms of the correlations via
the inclusion-exclusion principle (\ref{inclus}) and thus in terms of a
grand-canonical partition function $\Xi_\Lambda(z,V^{(1)},V^{(2)},V^{(3)})$
for a particle system in $\Lambda$ with fugacity $z$, one particle potential
$V^{(1)}$, non-translation-invariant pair potential $V^{(2)}$, and
translation invariant triplet potential $V^{(3)}$:
\begin{eqnarray}
p^\Lambda_n (\r_1,\ldots,\r_n) &=&  \rho^n \prod_{1\le i<j\le n}g(\r_j-\r_i)
   \prod_{1\le i<j<k\le n} \tilde g_3(\r_j-\r_i,\r_k-\r_i)\nonumber\\
 &&\hskip20pt\times\;
   \Xi_\Lambda(-\rho,V^{(1)},V^{(2)},V^{(3)}),
\end{eqnarray}
where
 \begin{eqnarray}\label{trippots3}
 V^{(3)}(\x,\y,\z) &:=& -\ln(\tilde{g}_3(\y-\x,\z-\x)), \\
 \label{trippots2}
 V^{(2)}(\x,\y) &:=& -\ln(g(\y-\x)) + \sum_{1\le i\le n}V^{(3)}(\x,\y,\r_i),\\
 \label{trippots1}
 V^{(1)}(\x)&:=& \sum_{1\le i\le n} V^{(2)}(\x-\r_i)
      + \sum_{1\le i<j\le n } V^{(3)}(\x,\r_i,\r_j).
\end{eqnarray}
 In order to proceed as in section~\ref{lowrho} we have to show that there
exists a domain for $z$, independent of $\r_1,\ldots,\r_n$, in which
$\ln\Xi_\Lambda(z, V^{(1)},V^{(2)},V^{(3)})$ is analytic.
Now, however, we must work with the cluster expansion for both the pair and
the triplet interactions, as in \cite{Gr71}.  We give a result in which the
hypotheses have been chosen to keep the proof simple and are thus far from
optimal.  We let $v_d$ denote the volume of the sphere in $\RR^d$ of
diameter $1$ and write
 \begin{equation}\label{C3def}
 C_3(\tilde g_3) = \sup_{\x,\y\in\RR^d}\max
  \{\,|\tilde{g}_3(\x,\y)-1|,|\tilde g_3(\x,\y)-1|^{1/3}\,\}.
 \end{equation}

 \begin{proposition} \label{probtriplet} Let $g$ and $\tilde{g}_3$ be be
given, and assume that (i) $g$ satisfies the conditions of
Remark~\ref{armremark}(c), and $C(g)<\infty$; (ii) there exists a $D_3>0$
such that $\tilde{g}_3(\x_2-\x_1,\x_3-\x_1)=1$ if $|\x_i-\x_j|> D_3$ for some
$i,j,k$, and $C_3(\tilde{g}_3)<\infty$.  Then $(\rho,g,\tilde{g}_3)$ is
realizable whenever
 \begin{equation} \label{eqcondtrip}
 0 \leq \rho \leq \left[ eb b_3
  \left(1+ bC_3(\tilde{g}_3) \right)^{(3D_3/D)^{2d}}
  \left(C(g)+ v_d (D_3/2)^d C_3(\tilde{g}_3)\right) \right]^{-1},
 \end{equation}
 where $b$ is defined as in Theorem~\ref{ThArmenianCluster} and $b_3$ is a
constant such that
\begin{equation}
  \prod_{j=1}^n\prod_{i=j+1}^n \tilde{g}_3(\x_i-\x_0,\x_j-\x_0) \le b_3
 \end{equation}
 for all $n$ and all $\x_0,\x_1,\ldots,\x_n$ with $|\x_i-\x_j|>D$ for
 $0\le i<j\le n$. .
\end{proposition}

\begin{remark}\label{tripremark}
 (a) Without loss of generality we may assume that $D_3>D$,
since otherwise $\tilde{g}_3=1$
and the proposition reduces to  Theorem~\ref{ThArmenianCluster}.

\noindent
(b) Since there can be at most
 \begin{equation}\label{Ndef}
 N := \left(\frac{2D_3+D}{D}\right)^d\leq \left(\frac{3D_3}{D}\right)^d
 \end{equation}
 points within a distance $D_3$ of $x_0$, all separated from $x_0$ and from
 each other by a distance at least $D$, we may in particular take
 \begin{equation}
  b_3 =    (1+C_3(\tilde{g}_3))^{(3D_3/D)^{2d}}. \nonumber
  \end{equation}
Note that with this choice the upper bound of (\ref{eqcondtrip}) converges
as $C_3(\tilde{g}_3) \searrow 0$ to the upper bound of
(\ref{ThArmeniancluster Req}).
 \end{remark}

\begin{proof}[Proof of Theorem \ref{probtriplet}:] The proof is a fairly
straightforward extension of the proof of Theorem~\ref{ThArmenianCluster}.
We sketch some details in appendix~\ref{armproof}.
\end{proof}

Finally we would like to point out a consequence of
Proposition~\ref{probtriplet} which illustrates the fact that the same
(finite) family of correlation functions may be realized by distinct
measures, and in fact by mutually singular measures, where two measures are
called {\sl mutually singular} if configurations of points typical for one
of the realizing point process are atypical for the other one, i.e., if
there exists a set of point configurations $A$ such that $A$ has
probability $1$ for one measure and probability $0$ for the other.

\begin{corollary} \label{nonuni} Let $g$ be a function fulfilling the
conditions of Remark~\ref{armremark}(c). Then for any $\rho$ satisfying the
bound (\ref{ThArmeniancluster Req}) of Theorem~\ref{ThArmenianCluster} with
strict inequality there exist uncountably many distinct and in fact
mutually singular  realizations of
$(\rho,g)$ by point processes.  \end{corollary}

\begin{proof}
  Under the hypotheses  one may choose $\tilde g_3$ quite
arbitrarily, subject only to a condition that $C_3(\tilde g_3)$ be
sufficiently small (how small depends on $D_3$), and still have $\rho$
satisfy the bound (\ref{eqcondtrip}).  Thus there are
certainly uncountably many realizations with distinct three point
functions.  To show that these are mutually singular
one first establishes, following the procedure of
subsection~\ref{decay}, the decay of the truncated correlation functions
for the measure constructed in Proposition~\ref{probtriplet}.
A direct consequence of this
decay is that the corresponding realizing point process is mixing and
therefore ergodic.  Since any two translation invariant ergodic measures
are either identical or mutually singular, the result follows.
\end{proof}

\section{The Lee-Yang approach} \label{subsecLee}

While for given $\rho_j$ we cannot in general
improve the region of realizability beyond that described in
section~\ref{lowrho} and \ref{triplet}, there are special situations in
which more can be said. One class of examples is treated in
appendix~\ref{examples}. In this section we consider a lattice gas on a
countable set $X$, e.g., $X=\ZZ^d$, with $G_2(\x,\y)\geq 1$ for
$\x,\y\in X$, $\x\ne\y$.  This enables us to to establish realizability
using techniques developed for proving the Lee-Yang theorem \cite{LY}.

\begin{theorem}
\label{theleeyang} Let $X$ be a countable set and suppose that
 $G_2(\x,\y) \geq 1$ for all
$\x,\y \in X$ with $\x \neq \y$ and that
 \begin{equation}\label{defB}
 b := \sup_{\x\in X}\prod_{\y \in X \setminus \{\x\}}
   G_2(\x,\y) < \infty.
 \end{equation}
Then $(\rho, G_2)$ is realizable for all
\begin{equation}
\label{ineqLeeYang} 0 \leq \rho \leq b^{-1}.
\end{equation}

\end{theorem}

The condition (\ref{ineqLeeYang}) improves the result of
Theorem~\ref{armenian-lat}, increasing the upper bound on $\rho$ by a
factor of $eC(g)$, with $C(g)=\sum_{\x\in \mathbb{Z}^d} |g(x)-1| $ as in
that theorem. Note that $C(g)\ge1$ since $g(0)=0$.

\begin{proof} We again use the ansatz (\ref{armansatz}) for the higher
correlation functions.  Let $\Lambda$ be a finite subset of $X$; then
(\ref{defB}) implies that $\rho_n$ fulfills the bound (\ref{clam}) with
$M_\Lambda= \rho b^{1/2}$. Hence we can write the probability densities
$p_n^\Lambda$ in terms of the correlation functions via (\ref{inclus}).
Since the
correlation function $\rho_{n+k}(\r_1,\ldots,\r_n,\x_1,\ldots,\x_k)$
vanishes when any of its arguments coincide, one can work
with the variables $\xi := \{ \r_1,\ldots, \r_n\}$ and
$\gamma = \{\x_1,\ldots,\x_k\}$, where $\xi$ and $\gamma$ vary over all
finite subsets of $\Lambda$ with $\gamma \cap \xi = \emptyset$.  Then
(\ref{inclus}) may be written in terms of $\gamma$ and $\xi$ as
$p^\Lambda_n(\xi) = \rho_{n}\!(\xi)\Xi^\Lambda_n(\xi)$,
where
 \begin{equation} \label{LeeYangbase}
 \Xi^\Lambda_n(\xi) =
     \sum_{\gamma \subset \Lambda \setminus \xi} (-\rho)^{|\gamma|}
    \prod_{\stackrel{ \scriptstyle \y \in \gamma}
   { \x \in \gamma \setminus\{\y\}} } G_2(\x,\y)^{1/2}
   \prod_{\stackrel{\scriptstyle \r \in \xi}{ \y \in\gamma}} G_2(\r,\y)
 \end{equation}
  (compare (\ref{partfunct})--(\ref{GCPF})).  As before the main problem is to verify that
$p^{\Lambda}_n \geq 0$. To apply techniques used for the Lee-Yang theorem
we write $p^{\Lambda}_n$ in terms of the set
$\sigma= (\Lambda \setminus \xi ) \setminus \gamma$ of empty sites rather
than in terms of $\gamma$. Writing $\tilde{\Lambda}:=\Lambda \setminus \xi$
we obtain
 \begin{eqnarray}
\Xi^\Lambda_n(\xi) &=& (-\rho)^{|\tilde{\Lambda} |}
\prod_{\y \in \tilde{\Lambda}} \left( \prod_{\r \in \xi} G_2(\r,\y)\
\hspace{-0.5cm} \prod_{\x \in \tilde{\Lambda} \setminus \{\y\}}
G_2(\x,\y)^{1/2} \right)\label{eqLeeYang}\\
 && \hspace{-1.5cm}\cdot
\sum_{\sigma \subset \tilde{\Lambda} }
\prod_{\y \in \sigma} \left( -\rho^{-1}\prod_{\r \in \xi} G_2(\r,\y)^{-1}
\hspace{-0.3cm}
\prod_{\x \in \tilde{\Lambda} \setminus \{\y\}}
G_2(\x,\y)^{-1/2}
\prod_{\x \in \tilde{\Lambda} \setminus \sigma }
G_2(\x,\y)^{-1/2}\right). \nonumber
 \end{eqnarray}
 Clearly the prefactor here is non-negative in general and is positive for
$\rho>0$.  To prove that the sum is non-negative we rewrite it in the form
 \begin{equation} \label{LeeYangform}
 \sum_{\sigma \subset
 \tilde{\Lambda}} \prod_{\y \in \sigma} \left(z_{\y} \prod_{\x \in
 \tilde{\Lambda} \setminus \sigma} A_{\r,\y} \right),
  \end{equation}
 where
 \begin{eqnarray}
 z_{\y} &:=& -\rho^{-1} \prod_{\r \in \xi} G_2(\r,\y)^{-1} \prod_{\x
\in \tilde\Lambda \setminus  \{\y\})} G_2(\x,\y)^{-1/2},\\
 A_{\x,\y}&:=& G_2(\x,\y)^{-1/2}.
  \end{eqnarray}
 Note that from $G_2(\x,\y)\geq 1$ for $\x \neq \y$ it follows that
$-1\leq A_{\x,\y} \leq 1$.  Then Proposition~5.1.1. of \cite{Ruelle}
implies that (\ref{LeeYangform}) is not zero if $|z_{\y}| > 1$ for all
$\y \in \Lambda \setminus \xi$. $|z_{\y}|$ can be bounded below by
$\rho^{-1}\prod_{\x
\in \Lambda \setminus \{\y\}} G_2(\x,\y)^{-1} \geq (\rho
b)^{-1}$.  We have thus shown that $\Xi^\Lambda_n(\xi)$ has no zeros for
$0 < \rho < 1/b$.  But $\Xi_n^{\Lambda}(\xi)=1$ for $\rho =0$, so that
$\Xi_n^\Lambda(\xi)$ and   hence
$p_n^\Lambda(\xi)$
is non-negative for all $0 \leq \rho \leq 1/b$.

\end{proof}

\section{Gibbsian measures \label{gibbs}}

In this section we ask whether a specified set of correlation functions
$\rho_j$, $j=1,\ldots,k$, which can be realized by at least one point
process, can also be realized by a Gibbs measure involving at most
$k$-particle potentials.  Here we will first consider this problem for the
case in which our system lives on a finite set $\Lambda$, e.g., a subset of
the lattice.  On a finite set every measure is Gibbsian in a general sense,
so the important restriction is to be Gibbsian for a set of potentials
involving at most $k$ particles: we will say that a measure $\nu$ on
$\{0,1\}^\Lambda$ is {\it $k$-Gibbsian} if it has the form
 \begin{eqnarray}\label{kGibbs}
  \nu(\eta)=Z^{-1}\exp\left
   \{-\sum_{j=1}^k\sum_{\x_1\ne\x_2\ne\cdots\ne\x_j\in\Lambda}
  \phi^{(j)}(\x_1,\ldots,\x_j)\eta(\x_1)\cdots\eta(\x_j)\right\},
 \end{eqnarray}
 where $\eta\in\{0,1\}^\Lambda$, $Z$
is a normalization constant, and  $-\infty<\phi^{(j)}\le\infty$.

As in (\ref{ntig}) we write, for $j=2,\ldots,k$,
 \begin{equation}
\rho_j(\x_1,\ldots,\x_j)=\prod_{i=1}^j\rho_1(\x_i)\;G_j(\x_1,\dots,\x_j),\qquad
 \x_1,\ldots,\x_j\in\Lambda,\label{gnti}
 \end{equation}
 and again think in terms of specifying the $G_j$, $j=2,\ldots k$, and
asking for what densities $\rho_1(\x)$ the correlations (\ref{gnti}) may be
realized by a $k$-Gibbs measure.  We will prove that this is possible
whenever $\rho_1(\x)$ satisfies $\rho_1(\x)<\bar\rho_1(\x)$ for
all $\x\in\Lambda$, with equality allowed if $\bar\rho_1(\x)=0$, for some
$\bar\rho_1$  with the property  that $\bar\rho_1$ and
the $G_j$, $j=2,\ldots,k$, are realizable.  The proof is presented only for
$k=2$, but the result could easily be extended to general $k$.

 The key ingredient in the argument is the fact that Gibbs measures are
those which maximize the Gibbs-Shannon entropy of the measure $\mu$,
\begin{equation}
S(\mu) \equiv -
 \sum_{\etaul} \mu(\etaul)\log \mu(\etaul)
\label{GS}
\end{equation}
  subject to some specified constraints \cite{Ruelle}.  In particular, if
one can use the method of Lagrange multipliers to find a measure which
maximizes the entropy, subject to the constraint of a given $\rho_1$ and
$\rho_2$, then the maximizing measure will be 2-Gibbsian and the Lagrange
multipliers obtained in this way will be the desired one body and pair
potentials \cite{Var}.  Here we verify that if $\rho_1(\x)<\bar\rho_1(\x)$
(with equality allowed if $\bar\rho_1(\x)=0$, as described above) then the
method of Lagrange multipliers will indeed apply.

\begin{theorem}\label{Gibbsmeasure} Suppose that the pair
$(\bar\rho_1,G_2)$ is realizable on $\Lambda$. If $\rho_1$ satisfies
$0\le\rho_1(\x)\le\bar\rho_1(\x)$ for all $\x\in\Lambda$, with
$\rho_1(\x)<\bar\rho_1(\x)$ unless $\bar\rho_1(\x)=0$, then $(\rho_1,G_2)$
is realizable by a 2-Gibbsian measure $\nu$ for some uniquely determined
potentials $\phi^{(1)}(\x)$, $\x\in\Lambda$ and $\phi^{(2)}(\x,\y)$,
$\x,\y\in\Lambda,\x\ne \y$.  Moreover, $\nu$ maximizes the Gibbs-Shannon
entropy (\ref{GS}) over all measures $\mu$ realizing $(\rho_1,G_2)$.
\end{theorem}

\begin{proof} We begin with a preliminary remark.  Suppose that we have
verified the theorem in the case in which all $\rho_1(\x)$ (and hence also
all $\bar\rho_1(\x)$) are strictly positive.  Then the case in which, say,
$\rho_1(\x)=0$ for $\x\in\Lambda'\subset\Lambda$, is a direct corollary: we
obtain immediately a 2-Gibbsian measure on
$\{0,1\}^{\Lambda\setminus\Lambda'}$ realizing the correlations there, and
then take $\phi^{(1)}(\x)=\infty$ for $\x\in\Lambda'$.  Similarly, if
$G_2(\x,\y)=0$ for some pair of sites $\x,\y\in\Lambda$, $\x\ne \y$, with
$\rho_1(\x)$ and $\rho_1(\y)$ nonzero, then we set
$\phi^{(2)}(\x,\y)=\infty$, which guarantees that if $\nu$ is given by
(\ref{kGibbs}) then $\nu(\eta)=0$ whenever $\eta(\x)=\eta(\y)=1$.  Thus in
the remainder of the proof we will assume that $\rho_1(\x)>0$ for all
$\x\in\Lambda$ and prove the existence of a realizing measure $\nu$, of the
form (\ref{kGibbs}) with {\it finite} potentials and with $k=2$, on the set
of configurations
 \begin{equation}\label{etacond}
 \mathcal{C}_{G_2}
   :=\left\{ \eta \mid
      \hbox{$\eta(\x)\eta(\y)=0$ if $G_2(\x,\y)=0$, $\x\ne \y$} \right\}.
 \end{equation}

 We now turn to the main body of the proof. For any $\eta$ we let
$|\eta|=\sum_{\x\in\Lambda}\eta(\x)$ be the number of particles in the
configuration $\eta$.  We first show that there exists a measure $\mu^*$
realizing $(\rho_1,G_2)$ for which $\mu^*(\eta)>0$ whenever
$\eta\in \mathcal{C}_{G_2}$ and $|\eta|\le2$. By hypothesis there
exists a measure $\bar\mu$ realizing $(\bar\rho_1,G_2)$.  We may thin this
measure as in section~\ref{intro}, deleting a particle at the site $\x$
with probability $1-\rho_1(\x)/\bar\rho_1(\x)$, independently for each
site, to obtain a measure $\mu^*$ realizing $(\rho_1,G_2)$. Now we observe
that if $\eta \in \mathcal{C}_{G_2}$ and $|\eta|\le2$ then
$\mu^*(\eta)>0$.  For example, if $|\eta|=2$ with $\eta(\x)=\eta(\y)=1$ for
$\x\ne\y$, then $G_2(\x,\y)>0$ by (\ref{etacond}) and hence, since
$\bar\mu$ realizes $(\bar\rho_1,G_2)$, $\bar\mu(\tilde\eta)>0$ at least for
one $\tilde\eta$ with $\tilde\eta(\x)=\tilde\eta(\y)=1$, and there is a
positive probability that $\eta$ will result from $\tilde\eta$ applying the
thinning process.

Next we construct a measure $\hat\mu$ realizing $(\rho_1,G_2)$ for which
$\hat\mu(\eta)>0$ for all $\eta \in \mathcal{C}_{G_2}$. We first fix
$\epsilon>0$ and for $|\eta|>2$ define
$\hat\mu(\eta)=\mu^*(\eta)+\epsilon$.  Now the condition that
$\hat\mu$ realize $(\rho_1,G_2)$ is that
 \begin{eqnarray}
\sum_{\eta\in\mathcal{C}_{G_2}} \hat\mu(\eta)&=&1, \label{constraint0}\\
\sum_{\eta\in\mathcal{C}_{G_2}} \eta(\x)\hat\mu(\eta)
    &=&\rho_1(\x),\qquad \x\in\Lambda,\label{constraint1}\\
\sum_{\eta\in\mathcal{C}_{G_2}} \eta(\x)\eta(\y)\hat\mu(\eta)
    &=&\rho_1(\x)\rho_1(\y)G_2(\x,\y),\qquad
   \x,\y\in\Lambda, \x\ne \y.  \label{constraint2}
 \end{eqnarray}
 Equations (\ref{constraint0})--(\ref{constraint2}) may be regarded as a
system of linear equations for the (as yet) undefined $\hat\mu(\eta)$,
$|\eta|\le2$; note that the number of these unknowns is the same as the
number of equations.  The coefficient matrix in this system is (after an
appropriate ordering of the $\eta$, $|\eta|<2$) upper triangular, with unit
diagonal; thus these equations can be solved uniquely for the
$\hat\mu(\eta)$, $|\eta|\le 2$, in terms of these given $\hat\mu(\eta)$,
$|\eta|>2$. The resulting $\hat\mu$ will differ from $\mu^*$
by a  perturbation of
order $\epsilon$; in particular, since $\mu^*(\eta)>0$ for
$|\eta|\le 2$, we can by choice of $\epsilon$ guarantee that also
$\hat\mu(\eta)>0$ for $|\eta|\le 2$.  But $\hat\mu(\eta)>0$ for $|\eta|>2$
by construction, so that $\hat\mu$ has the desired properties.

Finally we show that the Gibbsian measure we seek is the measure which
maximizes $S(\mu)$ (see (\ref{GS})) among all measures realizing
$(\rho_1,G_2)$.  Let $\nu$ be such a maximizer; $\nu$ is unique by the strict
concavity of $S$.  We first observe that $\nu$ must lie in the interior of
$\mathcal{C}_{G_2}$, i.e., that $\nu(\eta)>0$ for all
$\eta\in\mathcal{C}_{G_2}$; otherwise define
$\nu_t=(1-t)\nu+t\hat\nu$ and note that then
 \begin{equation}
\frac{d}{dt}S(\nu_t)\bigg|_{t=0}
 =-\sum_{\eta\in\mathcal{C}_{G_2}}\log\nu(\eta)=\infty,
 \end{equation}
 so that $S(\nu_t)>S(\nu)$ for some $t>0$ and $\nu$ cannot
maximize $S$.  Hence $\nu$ may be obtained by the method of Lagrange
multipliers, with (\ref{constraint0})--(\ref{constraint2}) (written for
$\nu$ rather than $\hat\mu$) as constraints.  A simple computation shows
that the $\nu(\eta)$ then have the form (\ref{kGibbs}) with $k=2$, where
the $\phi^{(1)}(\x)$ and $\phi^{(2)}(\x,\y)$ are the Lagrange multipliers
associated with (\ref{constraint1}) and (\ref{constraint2}), respectively
(the multiplier for (\ref{constraint0}) is related to the factor $Z$).

To verify uniqueness of the potentials, note that any 2-Gibbsian measure
realizing $(\rho_1,G_2)$ with, say, potentials $\psi^{(1)}$ and
$\psi^{(2)}$, must satisfy the Lagrange multiplier equations with these
potentials as multipliers and is hence an extremum of the entropy. From the
uniqueness of the extremum and the non-degeneracy of the constraint
equations (\ref{constraint0})--(\ref{constraint2}) it follows that these
multipliers are uniquely defined, i.e., that $\psi^{(1)}=\phi^{(1)}$ and
$\psi^{(2)}=\phi^{(2)}$.  \end{proof}

\subsection {Infinite volume \label{infvol}}

It is natural to ask if there exists an analogue of
Theorem~\ref{Gibbsmeasure} for an infinite lattice, such as $\ZZ^d$,
for example when the given correlation functions are defined from the beginning
on $\ZZ^d$ as translation invariant quantities, i.e., $\rho_1(\x)=\rho$,
$G_2(\x,\y)=g(\y-\x)$ as in (\ref{rho}), (\ref{g}); on $\ZZ^d$ we will call a
measure 2-Gibbsian if it satisfies the DLR equations for an interaction
with only one and two body potentials.  A result in this direction is due
to L.~Koralov \cite{Kor};  using cluster expansion techniques, he
has established the existence of an
infinite-volume 2-Gibbsian measure in the lattice case for $k=2$, $\rho$
small, and $g$ sufficiently close to $1$---specifically, for
$\sum_{\r \ne 0} |g(\r)-1| \leq 1$.

An attractive alternative approach would be to apply
Theorem~\ref{Gibbsmeasure} in large boxes $\Lambda\subset\ZZ^d$ to obtain
potentials $\phi^{(1)}_\Lambda(\x)$ and $\phi^{(2)}_\Lambda(\x,\y)$,
$\x,\y\in\Lambda$, realizing $(\rho,g)$ in $\Lambda$, and then to show that
under suitable restrictions on $\rho$ and $g$ the $\Lambda\nearrow\ZZ^d$
limits of these potentials exist and are summable and translation
invariant.  We do not know how to carry out such a program, as we have no
control on the behavior of the $\phi_\Lambda$ as $\Lambda$ changes.  In
fact we do not know if the 2-Gibbsian measures $\nu_\Lambda$ realizing
$(\rho,g)$ in $\Lambda$ converge as $\Lambda\nearrow\ZZ^d$ to any measure
$\nu$ on $\ZZ^d$.  Any sequence of such measures must have a
(weakly) convergent subsequence, however, by compactness, and the limiting
measure $\nu$ will realize the translation invariant $(\rho,g)$ on $\ZZ^d$.
In \cite{CKLS} we showed that any such realizing measure which is
translation invariant and Gibbsian, with summable potentials, is
necessarily a 2-Gibbsian measure with uniquely determined potentials which
maximizes the entropy density among all realizing translation invariant
measures.

 It follows from the above that if there is in fact a translation invariant
entropy maximizing 2-Gibbsian measure $\nu$ on $\ZZ^d$, realizing
$(\rho,g)$, with summable translation invariant potentials $\psi^{(1)}$ and
$\psi^{(2)}(\x-\y)$ ($\psi^{(1)}$ is just the chemical potential), then the
conditional measure $\nu(\cdot|\eta_{\Lambda^c})$ on $\Lambda$, for a
specified configuration $\eta_{\Lambda^c}$ on $\ZZ^d\setminus\Lambda$, will
be the 2-Gibbsian measure which maximizes the entropy for
$(\rho_1(x|\eta_{\Lambda^c}),G_2(x,y|\eta_{\Lambda^c}))$, the one and two
particle distributions obtained from $\nu(\eta_\Lambda|\eta_{\Lambda^c})$.
If furthermore this measure is unique for these potentials, then clearly
$\nu(\cdot|\eta_{\Lambda^c})\to\nu$ for every $\eta_{\Lambda^c}$ as
$\Lambda\nearrow\ZZ^d$.

To obtain translation invariant measures we may take for the domain
$\Lambda$ in Theorem \ref{Gibbsmeasure} a periodic lattice $\LL$, where
$\LL=\{-L+1,\ldots,L\}^d$ with periodic boundary conditions, and find a
measure $\nu_{\LL}$ realizing $(\rho,\rho g_\LL)$ for some periodic
$g_\LL(\r)$ defined for those $\r$ satisfying $|r_i|\le L$, $i=1,\ldots,d$,
and such that $g_\LL({\bf r}) \to g({\bf r})$ as $L \nearrow \infty$.  In
this case any subsequence limit $\nu$ of $\nu_{\LL}$ will be translation
invariant in $\ZZ^d$ and realize $(\rho,\rho g)$.
Conversely, from a translation
invariant $\mu$ we can construct $\mu_{\LL}$ by first projecting $\mu$ into
a cubical box $\Lambda$ of side $2L$ to obtain $\mu_\Lambda$, then
defining, for $\eta$ a configuration in $\Lambda$ or equivalently $\LL$,
$\mu_{\LL}(\eta)= (2L)^{-d}\sum_{\x\in\LL} \mu(\tau_\x\eta)$, where $\tau$
is the shift operator on $\LL$.  This yields a periodic measure $\mu_\LL$
with density $\rho$ and with $\tilde g_\LL(\r) = g(\r) + O(1/L)$ for fixed
$\r$.

The real question then is whether any subsequence limit of the $\nu_\LL$
will be a 2-Gibbs measure with summable pair potentials.  To answer this
requires some control of the potentials $\phi_\Lambda^{(2)}(\x-\y)$, which we
lack at present (see question 4 at section \ref{conclude}).

\section{Concluding remarks \label{conclude}}

There are clearly many natural questions left unanswered by our results.
We list some of them for the case of a specified $\rho$ and $g(\r)$.

1) Is there any practical way to bridge the gap between the obvious
necessary conditions described in section~\ref{secneccond} and the
sufficiency conditions given in sections \ref{lowrho} and \ref{subsecLee}?

2) When is the measure defined by (\ref{armansatz}) Gibbsian or
quasi-Gibbsian?

3) Can one extend Theorem {\ref{Gibbsmeasure}} to continuum systems in a
finite domain $\Lambda \subset \RR^d$?  We expect this to be true under
some reasonable assumptions on $g(\r)$, e.g., the hard core condition that
$g(\r) = 0$ for $r < D$, $D > 0$, under which there can only be a finite
number of particles in $\Lambda$.

4) What can one say about the existence and nature of an entropy maximizing
measure on $\ZZ^d$ for $(\rho,g)$?  In particular are there situations for
$\rho<\bar\rho$ when such a measure is not a translation invariant
2-Gibbsian measure?  As pointed out at the end of section \ref{gibbs}, if
the answer to this question is no then the $\nu$ obtained from the periodic
$\nu_\LL$ of formula (\ref{kGibbs}) will be the 2-Gibbs entropy maximizing
measure on $\ZZ^d$.

5) What happens to the realizability problem if one does not specify
$g(\r)$ for all $\r\in\ZZ^d$ (or $\RR^d$), but only for $\r$ in some finite
domain, say $|\r|\le R$?  (This question was mentioned briefly in
section{\ref{intro}}.)  As the notation indicates, we are still considering
translation invariant correlations; we may in addition require that the
two-point correlations $\rho_2(\x,\y)$ of the realizing measure all be
translation invariant and approach $\rho^2$ as $|\x-\y|\to\infty$.  This is
the problem discussed in \cite{Kanter} for $\r\in\ZZ$, or just on a ring.
It turns out that, at least for the case $\r\in\ZZ$ and specified $(\rho,
g(1),\ldots,g(k))$, one can compute, via a finite number of operations,
whether the correlations are realizable.  On the $d$ dimensional periodic
lattice $\LL$, for general $d$, the entropy maximizing measure $\nu_\LL$ in
(\ref{kGibbs}) will now contain only a finite number of terms, so that the
transition to an entropy maximizing 2-Gibbs measure on $\ZZ^d$ with only
finite range potentials appears feasible. This will be described in a
separate publication.

\appendix

\section{Simple examples\label{examples}}

In this appendix we collect some realizability results for certain
concretely given $g$. Let us first mention a realizability problem
in $\RR^d$ which has been extensively studied by Torquato and
Stillinger \cite{ST,ST2}: to determine for which densities $\rho$
there exists a translation invariant point process on $\RR^d$ with
 \begin{equation}
 g(\r)=
   \begin{cases}
   0,& \text{if $|\r|\le1$,}\\
      1,& \text{if $|\r|>1$.}
   \end{cases}\label{gc}
 \end{equation}
 Condition (\ref{FT}) implies that $(\rho,g)$ can be only realized if
$\rho\le (v_d 2^d)^{-1} $, where $v_d$ is the volume of the ball
with diameter $1$ in $\RR^d$ ($v_1=1$, $v_2=\pi/4$, etc.).  In the
other direction, Theorem~\ref{ThArmenianCluster} implies that for
general $d$ these correlations are indeed realizable if $\rho\le
e^{-1}v_d^{-1}2^{-d}$. Thus the maximum density $\bar\rho(d)$ for
which $g$ is realizable satisfies
 \begin{equation} \label{lowup}
e^{-1}\le2^dv_d\bar\rho(d)\le1.
 \end{equation}
 Ideas introduced in recent work  on the radius of convergence for pure
hard-core systems \cite{fernandez} can be used to give a stronger version
of Theorem~\ref{ThArmenianCluster}, improving the lower bound in
(\ref{lowup}) beyond $e^{-1}$, e.g., for $d=1$ to
$0.4$ and for $d=2$ to $0.5107$ (numerically). In one dimension we can say
more: a simple construction of \cite{CL} shows realizability by a renewal
process if $\rho\le 1/e$ (it is also shown in \cite{CL} that $1/e$ is the
maximal density for which a renewal process can realize $g$). A more
complicated construction \cite{CKLS}, using hidden Markov processes, gives
realizability for all $\rho \le 0.395$, so that
$0.395\le\bar\rho(1)\le0.5$.  The gap between these upper and lower bounds
remains as a challenge to further rigorous analysis. Certain simulation
results \cite{ST2} have suggested that in low dimensions the process may in
fact have $\bar\rho(d)=2^{-d}v_d^{-1}$. However, we can show that this is
false in one dimension, i.e., that $\bar\rho(1)<1/2$; we discuss this
further below.

This continuum problem is, in dimension $d=1$, related to the
following lattice problem: for what densities $\rho$ can the second
correlation function $\rho^2g^{(\alpha)}$, where
 \begin{equation}
 g^{(\alpha)}(x)=
   \begin{cases}
   0,& \text{if $x=0$,}\\
     \alpha,& \text{if $|x|=1$,}\\
      1,& \text{if $|x|>1$,}
   \end{cases}\label{hnew}
 \end{equation}
 be realized by a point process on $\ZZ$?  From the remarks in
section~\ref{intro} we know that for fixed $\alpha$ the set of
realizable densities $\rho$ is an interval $[0,\bar{\rho}_\alpha]$
with $0<\bar{\rho}_\alpha\leq 1$. There is a superficial similarity
between the continuum problem (\ref{gc}) and the lattice problem
(\ref{hnew}) for $\alpha=0$, but there is also a deeper relation.
For suppose that $\eta_c$ is a point process in $\RR$ which realizes
(\ref{gc}) at density $\rho$. Then, for any $k\in\ZZ$, define
$\eta_k=N_{(k,k+1]}=\int_{k-1}^k\eta_c(x)\,dx$ (that is, $\eta_k$ is
the number of points of the process $\eta_c$ lying in the interval
$(k,k+1]$). Then $\eta_k$ has value 0 or 1,
$\langle\eta_k\rangle=\rho$, and for $j>0$,
 \begin{eqnarray}\label{neq1}
\langle\eta_k\eta_{k+j}\rangle\;=\;\langle\eta_0\eta_j\rangle
 &=& \rho^2\int_0^1dx\int_j^{j+1}dy\, g(y-x)\nonumber\\
 &=& \begin{cases}\ds\rho^2\int_0^1dx\int_{1+x}^2 dy = \rho^2/2,&
     \text{if $j=1$,}\\
 \ds\rho^2\int_0^1dx\int_j^{j+1}dy =\rho^2,& \text{if $j\ge2$.}\end{cases}
 \end{eqnarray}
  Thus $\eta$ solves the lattice problem with the same density $\rho$ and
with $\alpha=1/2$, from which $\bar\rho(1)\le\bar\rho_{1/2}$.  We will see
below that $\bar\rho_{1/2}=1/2$, so that the relation
$\bar\rho(1)\le\bar\rho_{1/2}$ is consistent with the possibility
$\bar\rho(1)=1/2$ discussed above.

We now discuss the lattice problem (\ref{hnew}) in some
detail, as an illustration of the difficulties to face in the general
situation. Clearly $\bar\rho_1=1$, since for each $\rho\in[0,1]$ the
Bernoulli or product measure $\nu_\rho$ realizes (\ref{hnew}); for
$\rho\ne0,1$ there are in fact uncountably many mutually
inequivalent realizing measures \cite{CKLS}, while for $\rho=0,1$ the
realization is unique. For other values of $\alpha$, upper bounds on
$\bar{\rho}_\alpha$ are provided by (\ref{FT}) and (\ref{Yamada}).
In particular, (\ref{FT}) yields the upper bound $\bar\rho_\alpha\le
R_F(\alpha)$, where
\begin{equation}
\label{Srhoalpha}
  R_F(\alpha) = \left\{ \begin{array}{ll}
    \ds\frac{1}{3-2\alpha},& \mbox{if } 1 \geq \alpha \geq 0, \\
    \ds\frac{1}{2 \alpha -1}, & \mbox{if } \alpha \geq1.
\end{array}\right.
\end{equation}
The Yamada condition (\ref{Yamada}) gives an upper bound
$\bar\rho_\alpha\le R_Y(\alpha)$; a straightforward but somewhat lengthy
computation shows that $R_Y(\alpha)=R_F(\alpha)$ for $\alpha=1/2$, for
$\alpha=(k\pm1)/2k$, $k=1,2,\ldots$, and for $\alpha\ge1$. For other values of
$\alpha$, $R_Y(\alpha)<R_F(\alpha)$, so that certainly the bound $R_F$ is
not always sharp.  These bounds, together with several lower bounds for
$\bar\rho_\alpha$ obtained below, are plotted in Figure~1; for values of
$\alpha$ at which $R_Y(\alpha)<R_F(\alpha)$, $R_Y(\alpha)$ is determined
numerically.

\begin{figure} [!t]
\centerline{\epsffile{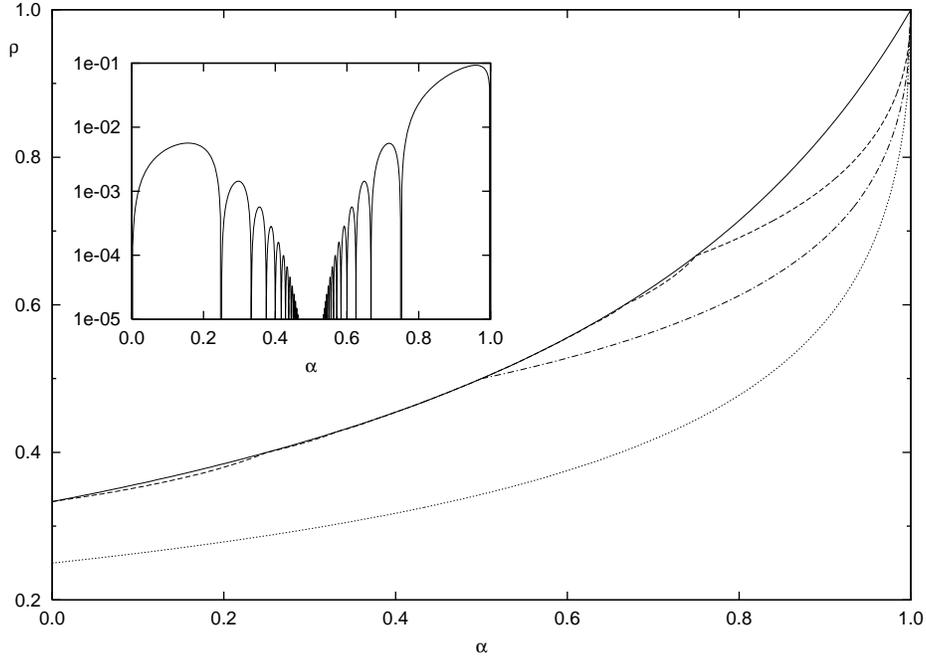}}

\caption{Upper bounds $R_F(\alpha)$ (solid) and $R_Y(\alpha)$
(dashes) for $\bar\rho_\alpha$. Lower bounds $r_S(\alpha)$
(dots/dashes) and $r_B(\alpha)$ (dots), for $0\le\alpha\le1$. The
inset plots the difference $R_F(\alpha)-R_Y(\alpha)$ on a
logarithmic scale.}

\end{figure}

Lower bounds on $\bar\rho_\alpha$ come, essentially, from procedures for
explicitly realizing the desired process at some value of $\rho$. For
example, from Theorem~\ref{ThArmenianCluster}, for $\alpha\le1$, and
Theorem~\ref{theleeyang}, for $\alpha\ge1$, we obtain
$\bar\rho_\alpha\ge r_A(\alpha)$, where
 \begin{equation}
r_A(\alpha) = \begin{cases}
  \ds\frac1{e (3-2\alpha)},& \hbox{if $1\ge\alpha\ge0$},\cr
  \ds\frac1{\alpha^2},& \hbox{if $\alpha\geq 1$}.\cr
 \end{cases}\label{genbds}
 \end{equation}
 Comparison with (\ref{Srhoalpha}) shows that, as might be expected, the
lower bounds from these general construction methods do not approach the
upper bound very closely. To get better bounds or exact values for
$\bar{\rho}_\alpha$ one must turn to more {\sl ad hoc} methods.  In this
spirit we next describe two families of processes which realize
(\ref{hnew}) and which provide improvements in the lower bounds
(\ref{genbds}).  We studied other constructions, partially improving some
of the results below, but these are omitted for conciseness.

The first construction, valid for $\alpha\ge1/2$, achieves a density
$\rho=r_S(\alpha)$, where
 \begin{equation}\label{super}
  r_S(\alpha) = \left\{ \begin{array}{ll}
    \ds\frac{1}{1+\sqrt{2-2\alpha}},
    & \mbox{if } 1 \geq \alpha \geq 1/2, \\
    \ds\frac{1}{2 \alpha -1}, & \mbox{if }  \alpha\ge1;
\end{array}\right.
   \end{equation}
 thus $\bar\rho_\alpha\ge r_S(\alpha)$.  Comparison of (\ref{super}) with
(\ref{Srhoalpha}) shows that $r_S(\alpha)=R_F(\alpha)$ for
$\alpha=1/2$ and $\alpha\ge1$, so that $\bar\rho_\alpha=R_F(\alpha)$
for these values. The measures for these processes are
superpositions of two measures of period two.  To construct them, we
first choose with equal probability one of two partitions of $\ZZ$,
either $\ldots\cup \{-2,-1\} \cup \{0,1\} \cup \{2,3\}\cup \ldots$
or $\ldots\cup \{-1,0\} \cup \{1,2\} \cup \{3,4\}\cup\ldots$, and
then assign a configuration to each pair $(i,i+1)$ of sites in the
partition independently, taking $(\eta_i,\eta_{i+1})$ to have value
$(1,0)$, $(0,1)$ each with probability $p$, $(0,0)$ with probability
$q$, and $(1,1)$ with probability $(1-p-q)/2$. The optimal choices
of parameters which lead to (\ref{super}) are $q=0$,
$p=\sqrt{2-2\alpha}/(1+\sqrt{2-2\alpha})$ for $1/2\le\alpha\le1$ and
$p=0$, $q=(2\alpha-2)/(2\alpha-1)$ for $\alpha \geq 1$.

The measures constructed above, as superpositions of period-two
measures, do not have good mixing properties (this defect will be
inherited by measures with lower values of $\rho$ obtained via the
thinning process described in section~\ref{intro}). For the case
$\alpha=\rho=1/2$ this decomposability is inevitable; the system is
then superhomogeneous (see section~\ref{secneccond}; in fact the
variance of the number of points on any set of consecutive lattice
sites is uniformly bounded by 3/4) and the decomposability of any
realizing measure then follows from a result of Aizenmann,
Goldstein, and Lebowitz \cite{AGL}.

We now sketch briefly the argument, mentioned in the first paragraph
of this appendix, that yields $\bar\rho(1)<1/2$.  Given a
point process $\eta_c$ in $\RR$ realizing (\ref{gc}) at density
$\rho$, we define a lattice process $\eta^{(n)}$, $n\ge1$, by
$\eta^{(n)}_k=\int_{(k-1)/n}^{k/n}\eta_c(x)\,dx$. Then $\eta^{(n)}$
has density $\rho/n$ and two point function
 \begin{equation}
 \langle\eta_0\eta_j\rangle=
   \begin{cases}
   0,& \text{if $|j|<n$,}\\
   \rho^2/(2n^2) ,& \text{if $|j|=n$,}\\
      \rho^2/n^2,& \text{if $|j|>n$.}
   \end{cases}\label{hn}
 \end{equation}
 For $n=1$ this is just the construction given earlier (see (\ref{neq1})),
and as there we see that if $\rho^{(n)}_*$ is the maximum density
for which (\ref{hn}) can be realized then necessarily
$\bar\rho(1)\le n\rho^{(n)}_*$; thus $\bar\rho(1)<1/2$ follows from
$\rho^{(2)}_*<1/4$. To verify the latter, one first shows that any
realizing measure $\mu_1$ for (\ref{hn}) with $n=1$, $\rho=1/2$ is
supported on configurations such that for some $i$ with $i=1,2$,
each pair of sites $(2j+i,2j+i+1)$, $j \in \ZZ$, contains exactly
one particle; this follows from the fact that the random variable
$\sum_{i=j}^{j+2k}\eta_i$ ($j \in \ZZ$, $k\in \NN$) has minimal
variance $1/4$ and hence takes only the values $k$ and $k+1$.  (Note
that these configurations form the support of the $\alpha = 1/2$
measure constructed above.)  Such a $\mu_1$ may be obtained by
projecting a realizing measure $\mu_2$ for (\ref{hn}) with $n=2$,
$\rho=1/2$ via either $\eta'_k := \eta^{(2)}_{2k}
+\eta^{(2)}_{2k+1}$ or $\eta''_k := \eta^{(2)}_{2k-1}
+\eta^{(2)}_{2k}$, and it then follows that $\mu_2$ is supported on
configurations with the property that for some $i$ with $1\le
i\le4$, all sites $4j+i$, $j\in\ZZ$, are empty and each triple of
sites $(4j+i+1,4j+i+2,4j+i+3)$ contains exactly one particle. This
leads to a decomposition $\mu_2 = \sum_{i=1}^4 \mu_{2,i}$ and from
this we derive a contradiction, since the variance with respect to
$\mu_{2,1}$ of one of the random variables $\sum_{k=1}^m
\eta^{(2)}_{4k+i}$, $i=1,\ldots,4$,  would be negative for large
enough $m$.  For details see \cite{KLS2}.

Our second construction is valid for $0\leq \alpha \leq 1$.  For this
process we first distribute particles on $\mathbb{Z}$ with a Bernoulli
measure such that each site is occupied with probability $\lambda$; then if
in this initial configuration a site is occupied we delete the particle
occupying its left neighbor, if it exists, with probability $\kappa$.  With
the optimal choices $\lambda=1/(1+\sqrt{1-\alpha})$ and
$\kappa=\sqrt{1-\alpha}$ we obtain a realization of (\ref{hnew}) with
density
 \begin{equation}\label{Bern}
  r_B(\alpha)=\frac{1}{(1+\sqrt{1-\alpha})^2},
 \end{equation}
 so that  $\bar{\rho}_\alpha\ge r_B(\alpha)$. For $\alpha\ge1/2$
this is of no interest, since $r_B\le r_S$, but for $0\le\alpha<1/2$
it improves on the bounds (\ref{genbds}).

The case $\alpha=0$ merits special discussion.  From (\ref{Bern}),
$\bar\rho_0\ge0.25$; in this case the point process used to obtain $r_B$ is
a renewal process and the construction is a lattice version of that given
in \cite{CL} to establish that $\bar{\rho}\ge1/e$ for the continuum problem
(\ref{gc}). A construction based on a hidden Markov process \cite{CKLS}
improves this lower bound to $\bar\rho_0 > 0.265$.  The upper bound
$\bar\rho_0\le R_F(\alpha)=R_Y(\alpha)=1/3$ can be improved \cite{CKLS} to
$\bar{\rho}_0\le(326-\sqrt{3115})/822\simeq0.3287$.  As in the continuum
problem (\ref{gc}), it remains a challenge to diminish the rather large gap
between these upper and lower bounds.

\section {Proofs of Theorems \ref{ThArmenianCluster} and \ref{probtriplet}
    \label{armproof}}

\begin{proof}[Completion of the proof of Theorem~\ref{ThArmenianCluster}]
We must show that the functions $k_m^\Lambda$ are analytic for $|z|<R$.
As indicated in section~\ref{lowrho}, the proof follows closely the proof
of Theorem 4.2.3 of \cite{Ruelle}, and we content ourselves with pointing
out a few key steps and the necessary changes.  Let $E_\xi$ be the Banach
space of all sequences $\left(\varphi_m \right)_{m=1}^\infty$, where
$\varphi_m:(\RR^d)^m\to\CC$, for which
 \begin{equation}\label{norm}
 \|\varphi\|_{\xi} :=\sup_{m\ge0}\sup_{\x_1,\ldots ,\x_m \in \RR^d}
  \xi^{-m}|\varphi_m(\x_1,\ldots,\x_m)| < \infty,
 \end{equation}
 and let $\chi_m(\x_1,\ldots,\x_m)$ be the characteristic function of
the set
 \begin{equation}\label{set}
 \biggl\{\,(\x_1,\ldots\x_m)\in(\RR^d)^m\,\bigg|\, \x_i\in\Lambda,\
\prod_{i=1}^m \left(\prod_{j=i+1}^m g(\x_i -\x_j) \prod_{j=1}^n
g(\x_i-\r_j) \right)>0\,\biggr\} .
 \end{equation}
 Define the  operator $\K$ on $\bigcup_{\xi>0}E_\xi$ by
\begin{eqnarray}\label{Kdef}
(\K\varphi)_{m+1}(\x_1,\ldots,\x_m,\x) &=&
  \chi_{m+1}(\x_1,\ldots,\x_m,\x)
e^{-V^{(1)}(\x)-\sum_{j=1}^m V^{(2)}(\x-\x_j)}
  \nonumber\\
 &&\hskip5pt
  \cdot\, \sum_{k=\max\{0,1-m\}}^\infty \frac{1}{k!} \int_{\RR^{dk}}
 \prod_{j=1}^k \left( e^{-V^{(2)}(\x- \z_j)}-1\right)
\nonumber\\
 &&\hskip5pt
  \cdot \, \varphi_{m+k}(\x_1,\ldots,\x_m,\z_1,\ldots,\z_k)
 \,d\z_1\cdots d\z_k.\qquad
\end{eqnarray}
 The factor $\chi_{m+1}$ in (\ref{Kdef}) is needed in the estimate
(\ref{Kest}) below because the one-body potential $V^{(1)}$ in
(\ref{Kdef}) depends on $\r_1,\ldots,\r_n$, and it is primarily these
aspects which distinguish the proof here from that of \cite{Ruelle}. Note
that, with this factor, $\K$ depends on $\Lambda$, on $g$, and on
$\r_1,\ldots,\r_n$.  Then
\begin{eqnarray}\label{Kest}
  |(\K\varphi)_{m+1}(\x_1,\ldots,\x_m,\x)| &\leq&
  \|\varphi\|_\xi  e^{-V^{(1)}(\x)-\sum_{j=1}^m V^{(2)}(\x-\x_j)}
       \nonumber \\
  &&\hskip-120pt\cdot\chi_{m+1}(\x_1,\ldots,\x_m,\x)
  \sum_{k=0}^\infty \frac{1}{k!}
  \int_{\RR^{dk}} \prod_{j=1}^k
  \left| e^{-V^{(2)}(\x- \z_j)}-1\right| \xi^{k+m}
   d\z_1\cdots d\z_k \nonumber \\
&&\hskip-130pt\le\;
    \| \varphi\|_{\xi}  \xi^m\  b \exp\left(\xi \int_{\RR^d} |
g(\z)-1| d\z\right),
\end{eqnarray}
 so that
 \begin{equation}
\|\K\varphi\|_\xi\le  \xi^{-1} b \exp\left(\xi \int_{\RR^d} |g(\z)-1|
d\z\right)
    \|\varphi\|_{\xi}.
 \end{equation}
  Now as in \cite{Ruelle} the sequence of correlation functions
$k^\Lambda_m$ satisfies the Kirkwood-Salsburg equation
 \begin{equation}\label{KS}
   k^\Lambda = z\psi+z\,\K k^\Lambda
 \end{equation}
 where $\psi_m=\delta_{m1}$; by an optimal choice of $\xi$ we may insure
 that $\|z\K\|_\xi<1$ whenever $|z|<R$. For such $z$, $I-z\K$ is invertible
 and the equation (\ref{KS}) has a unique solution in $E_\xi$; one
 proceeds as in \cite{Ruelle}.
\end{proof}

\begin{proof}[Proof of Theorem~\ref{probtriplet}] As in the proof of
Theorem~\ref{ThArmenianCluster} we denote by $k^\Lambda_m$ the $m^{\rm th}$
correlation function for a grand canonical ensemble in $\Lambda$ with
activity $z$, inverse temperature $\beta$, and potentials
(\ref{trippots3})--(\ref{trippots1}) ; we do not write explicitly the
dependence of these functions on $\r_1,\ldots,\r_n$.  The $k^\Lambda$ again
satisfy the Kirkwood-Salsburg equation (\ref{KS}) in the Banach space
$E_\xi$, but with a modified operator $\K$.  To define $\K$ we write
$\u_i:=\r_i$ for $i=1,\ldots,n$, $\u_i:=\x_{i-n}$ for $i=n+1,\ldots,n+m$,
and $\u_i:=\z_{i-n-m}$ for $i=n+m+1,\ldots,n+m+k$.  Then
 \begin{eqnarray}
(\K\varphi)_{m+1}(\x_1,\ldots,\x_m,\x) &=&
 ze^{-E(\x|\u_1,\ldots,\u_{n+m})} \sum_{k=0}^\infty
\frac{1}{k!}\int_{\Lambda^k}\label{kirkwood} \\
\nonumber
&&\hskip-140pt \cdot K(\x|\u_1,\ldots,\u_{n+m+k})
   \varphi_{m+k}(\u_{m+1},\ldots,\u_{n+m+k})\,
  d\u_{n+m+1}\cdots d\u_{n+m+k},
\end{eqnarray}
 with
\begin{eqnarray}\label{Edef}
E(\x|\u_1,\ldots,\u_{n+m}) &:=& \sum_{i=1}^{n+m} V^{(2)}(\x,\u_i) +
\hspace{-0.4cm} \sum_{1 \leq i < j \leq n+m} \hspace{-0.5cm}
V^{(3)}(\x,\u_i,\u_j) \\
  K(\x|\u_1,\ldots,\u_{n+m+k})
 &:=&   \chi_{n+m+k+1}(\u_1,\ldots,\u_{n+m+k},\x)\nonumber\\
 &&\hspace{-2.5cm}\cdot \sum_{\eta \subset
   \{1,\ldots,k \}} \prod_{i\in\{1,\ldots,k\}\setminus\eta}
   \left(e^{-V^{(2)}(\x,\u_{n+m+i})}-1 \right) \nonumber \\
&&\hspace{-2.5cm}\cdot \sum_{G } \prod_{\{i,j\} \in G}
  \left(e^{-V^{(3)}(\x,\u_i,\u_j)}-1\right) \prod_{i \in \eta}
   e^{-V^{(2)}(\x,\u_{n+m+i})}. \label{K3def}
\end{eqnarray}
 Here $\sum_G$ extends over the set of graphs which have vertex set
$V_1\cup V_2$, where $V_1=\{1,\ldots,m+n \}$ and
$V_2=\{l+m+n\mid l \in \eta \}$ (a graph being identified as a set of
edges, i.e., of unordered pairs of vertices), and in which every edge has
at least one of its vertices lying in $V_2$, and every vertex in $V_2$ has
at least one edge incident on it.  The derivation of
(\ref{kirkwood}--\ref{K3def}) is similar to that of  the usual
Kirkwood-Salsburg equation \cite{Ruelle}.

The operator $\K$ can be bounded as follows.  First, the hypotheses of the
theorem imply that $\chi_{n+m+k+1}(\u,\x)e^{-E(\x|\u)} \leq b\ b_3$.  Next,
since the factor $(e^{-V^{(3)}(\x,\u_i,\u_j)}-1)$ in (\ref{K3def}) vanishes
unless $\u_i$ and $\u_j$ lie inside the ball of radius $D_3$ around $\x$,
the sum over $\eta$ is nonzero only for  those $\eta$'s such that all
points $(\z_i)_{i\in \eta}$ are inside this ball. On the other hand, the
factor $\chi_{n+m+k+1}(\u,\x)$ implies that $K$ vanishes unless all the
$\u_i$ are separated by a distance at least $D$, so that we may suppose
that there are at most $N$ (see (\ref{Ndef})) of these points inside the
ball.   We thus have the bound
 \begin{eqnarray}
\lefteqn{\int_{\Lambda^k}|K(\x|\u_1,\ldots,\u_{n+m+k})|\,
  d\u_{n+m+1}\cdots d\u_{n+m+k}}
   \nonumber \hskip60pt &&\\
 &\leq& \sum_{l=0}^N 
 \binom kl
 C(g)^{k-l} (v_d (D_3/2)^d)^l \sum_{j=\lceil l/2\rceil}^M
 \binom Mj \hat C_3^j b^j
 \nonumber \\
&\leq& \left(1+ b C_3(\tilde{g}_3) \right)^M
  \left(C(g) + v_d(D_3/2)^dC_3(\tilde{g}_3) \right)^k.
\label{Best}
\end{eqnarray}
 Here $M=N(N-1)/2$,
$\hat C_3 = \sup_{\x,\y\in\RR^d} |\tilde{g}_3(\x,\y)-1|$, $\lceil s\rceil$
is the least integer not smaller than $s$, and we have used the
inequality $\hat C_3^j \leq C_3(\tilde g_3)^jC_3(\tilde g_3)^l$, valid for
$j\ge l/2$, which follows from (\ref{C3def}) by considering
separately the cases $\hat C_3\ge1$ and $\hat C_3<1$.

 From (\ref{Ndef}) we have that $M\le(3D_3/D)^{2d}$, and from
(\ref{kirkwood}) and (\ref{Best}) it then follows that the norm of
$\K$ in the Banach space $E_\xi$ satisfies
 \begin{equation}
  \|\K\|_\xi
  \le \xi^{-1} b b_3 \left(1+ b C(\tilde{g}_3) \right)^{(3D_3/D)^{2d}}
e^{\xi\left(C(g) + v_d (D_3/2)^dC(\tilde{g}_3) \right)}.
\end{equation}
 An optimal choice of $\xi$ again shows that $\|z\K\|_\xi<1$ when
$|z|<R_3$, where $R_3$ denotes the right hand side of (\ref{eqcondtrip}),
 completing the proof.
\end{proof}

{\bf Acknowledgments}: We thank E.~Caglioti, R.~Fernandez,
G.~Gallavotti, I.~Kanter, Yu.~G.~Kondratiev, P.~P.~Mitra,
J.~K.~Percus, F.~Stillinger, S.~Torquato, A.~van Enter, and
R.~Varadhan for valuable comments. We also thank the IHES, and
J.L.L.\ and E.R.S.\ thank the IAS for hospitality during the course
of this work. The work of T.K. was supported by the A.~v.~Humboldt
Foundation.  The work of J.L.L.\ and T.K. was supported by NSF Grant
DMR-0442066 and AFOSR Grant AF-FA9550-04.  We also thank DIMACS and
its supporting agencies. Any opinions, findings, conclusions, or
recommendations expressed in this material are those of the authors
and do not necessarily reflect the views of the National Science
Foundation.

\end{document}